\documentclass[sigconf]{aamas}

\usepackage{balance}
\usepackage{graphicx}
\usepackage{float}
\usepackage{placeins}
\usepackage[inline]{enumitem}
\usepackage{fancyvrb}
\usepackage{caption}
\usepackage{booktabs}

\makeatletter
\gdef\@copyrightpermission{
  \begin{minipage}{0.2\columnwidth}
   \href{https://creativecommons.org/licenses/by/4.0/}{\includegraphics[width=0.90\textwidth]{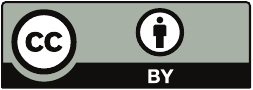}}
  \end{minipage}\hfill
  \begin{minipage}{0.8\columnwidth}
   \href{https://creativecommons.org/licenses/by/4.0/}{This work is licensed under a Creative Commons Attribution International 4.0 License.}
  \end{minipage}
  \vspace{5pt}
}
\makeatother

\setcopyright{rightsretained}
\acmConference[]{}
\copyrightyear{2026}
\acmYear{}
\acmDOI{}
\acmPrice{}
\acmISBN{}

\title[MOASEI Competition AAMAS'2026 Technical Report]{Second MOASEI Competition at AAMAS'2026: A Technical Report}

\author{Ceferino Patino}
\affiliation{
  \institution{University of Nebraska - Lincoln}
  \city{Lincoln}
  \state{Nebraska}
  \country{United States}}
\email{cpatino2@unl.edu}

\author{Tyler J. Billings}
\affiliation{
  \institution{University of Nebraska - Lincoln}
  \city{Lincoln}
  \state{Nebraska}
  \country{United States}}
\email{tbillings4@unl.edu}

\author{Alireza Saleh Abadi}
\affiliation{
  \institution{University of Nebraska - Lincoln}
  \city{Lincoln}
  \state{Nebraska}
  \country{United States}}
\email{asalehabadi2@unl.edu}

\author{Daniel Redder}
\affiliation{
  \institution{University of Georgia - Athens}
  \city{Athens}
  \state{Georgia}
  \country{United States}}
\email{daniel.redder@uga.edu}

\author{Adam Eck}
\affiliation{
  \institution{Oberlin College}
  \city{Oberlin}
  \state{Ohio}
  \country{United States}}
\email{aeck@oberlin.edu}

\author{Prashant Doshi}
\affiliation{
  \institution{University of Georgia - Athens}
  \city{Athens}
  \state{Georgia}
  \country{United States}}
\email{pdoshi@uga.edu}

\author{Leen-Kiat Soh}
\affiliation{
  \institution{University of Nebraska - Lincoln}
  \city{Lincoln}
  \state{Nebraska}
  \country{United States}}
\email{lksoh@unl.edu}

\begin{abstract} 
We describe the 2026 Methods for Open Agent Systems Evaluation Initiative (MOASEI) Competition, a benchmark event for evaluating multi-agent decision-making under open-system conditions. Building on the inaugural 2025 competition, the 2026 edition retained wildfire fighting, cybersecurity, and ride-sharing domains while adding a bonus wildfire track with frame openness, in which agent equipment states such as suppressant capacities and firefighting range vary over time. The competition also expanded its reporting metrics to emphasize total task completions, mean task-completion time, and mean value of completed tasks. Participation in 2026 was limited: eight teams registered, but only one team submitted a final entry, and that entry targeted the ride-sharing track. The submitted DLC approach used planning and replanning to solve routing problems across agents as passengers appeared. This report summarizes the 2026 competition design, highlights differences from the previous year, and reports ride-sharing evaluation results against baseline policies. DLC is recognized as the 2026 ride-sharing track winner among submitted teams.
\end{abstract}


\newcommand{\BibTeX}{\rm B\kern-.05em{\sc i\kern-.025em b}\kern-.08em\TeX}

\begin{document}

\pagestyle{fancy}
\fancyhead{}

\maketitle

\section{Introduction}

The second annual Methods for Open Agent Systems Evaluation Initiative (MOASEI) Competition was held on May 25, 2026 in Paphos, Cyprus as part of the 25th International Conference on Autonomous Agents and Multiagent Systems (AAMAS'2026). The MOASEI competition series evaluates multi-agent AI systems in open agent environments where agents, tasks, and the capabilities of agents may change over time. The competition is built on the free-range-zoo environment suite~\cite{free_range_zoo_repository}, which implements benchmark domains inspired by open agent systems~\cite{eck2023_decision}. The inaugural competition in 2025~\cite{patino2025_inaugural} established the core evaluation infrastructure and introduced three benchmark tracks: wildfire fighting, cybersecurity, and ride-sharing. 

The 2026 competition was designed to build upon the initial success of  MOASEI'2025.  Rather than introducing an entirely new benchmark suite, the goals were to refine the evaluation tracks, add a new form of openness, improve the interpretability of results through more task-centered metrics, and continue to grow the community of researchers addressing the real-world complexities imposed by open agent systems. Participation was lower than in the inaugural year: eight teams registered, one team submitted a final solution, and only the ride-sharing track received a submission. This report focuses on documenting the 2026 updates and providing a concise analysis of the winning ride-sharing submission relative to baseline policies.

\section{Background}

MOASEI focuses on decision making in open multi-agent environments. In this setting, the decision making of autonomous agents must handle changes that are not limited to stochastic state transitions. Agents may enter or leave the system, tasks may appear or disappear, and the capabilities or relevant state variables of agents may change over time. We refer to these dimensions as \emph{agent openness} (AO), \emph{task openness} (TO), and \emph{frame openness} (FO), respectively.

As in the previous competition~\cite{patino2025_inaugural}, each domain can be viewed as a partially observable stochastic game (POSG) with state, action, transition, reward, observation, and observation-function components. The 2026 edition continued the established wildfire fighting (focused on AO+TO), cybersecurity defense (focused on AO), and dynamic ride-sharing (focused on TO) settings, and also extended the wildfire fighting setting with a bonus track that explicitly adds frame openness. This report does not repeat the full domain formalizations from the 2025 technical report~\cite{patino2025_inaugural}; instead, it summarizes the operational differences and reports this year's results.

\subsection{Summary of the 2025 MOASEI Competition}

Here we provide a brief summary of the 2025 MOASEI competition to provide further contexts for our current MOASEI competition. For additional details on the 2025 MOASEI competition, please refer to that competitition's technical report~\cite{patino2025_inaugural}.

The inaugural MOASEI Competition at AAMAS 2025 established the initial benchmark structure and evaluation pipeline for open agent systems. It introduced three tracks built on free-range-zoo: wildfire fighting, cybersecurity defense, and dynamic ride-sharing. These tracks were designed to evaluate complementary forms of openness. Wildfire combined agent openness and task openness through changing agent availability and dynamically spreading or appearing fires. Cybersecurity emphasized agent openness through changing defender and attacker presence. Ride-sharing emphasized task openness through passengers that entered the system over time.  

The 2025 competition attracted eleven registered teams and four final submissions. Submitted approaches leveraged graph neural networks, convolutional neural networks, weighted heuristics, predictive modeling, and large language model (LLM)-assisted strategies. Final submissions were concentrated in the wildfire fighting and cybersecurity defense tracks, while the dynamic ride-sharing track received no final submissions. In wildfire fighting, the strongest submissions performed comparably to the best heuristic baseline, with Markov Mayhem and the University of Tehran recognized as track winners. In cybersecurity, Zana Cyber was recognized as the track winner for a weighted scoring approach that achieved strong rewards while using fewer patch actions and more monitoring than the baseline policies.

The main lessons from 2025 were that open agent system benchmarks can expose meaningful differences in coordination, robustness, and adaptation strategies, but that accessibility and track design strongly influence participation. The 2025 report therefore identified several directions for the next competition, including larger and more realistic environments, stronger attacker behavior in cybersecurity, a revised ride-sharing track with more direct competition, and new mechanisms for evaluating changes in agent capabilities. The 2026 competition follows directly from those lessons by adding a bonus wildfire track with frame openness and by emphasizing more interpretable task-level metrics.

\section{Overview of the MOASEI Competition}

This section summarizes the structure of the 2026 MOASEI Competition. We first describe the main changes from the 2025 competition, including the addition of frame openness and more task-centered evaluation metrics. We then provide a brief summary of registration and final submission activity.  

Table~\ref{tab:tracks_2026} summarizes the four evaluation tracks offered in 2026 and the forms of openness emphasized by each track.

\begin{table*}[htbp]
  \centering
  \caption{MOASEI 2026 evaluation tracks. AO denotes agent openness, TO denotes task openness, and FO denotes frame openness.}
  \begin{tabular}{p{0.22\textwidth}p{0.12\textwidth}p{0.58\textwidth}}
    \toprule
    \textbf{Track}          & \textbf{Openness} & \textbf{Description}                                                                                                            \\
    \midrule
    Wildfire Fighting       & AO/TO             & Agents collaborate with limited resources to fight fires of varying sizes. Fires spread between cells and may randomly ignite.  \\
    Bonus Wildfire Fighting & AO/TO/FO          & Agents have changing suppressant capacities and firefighting ranges, influencing both their duration in the environment and the tasks they can reach.    \\
    Cybersecurity           & AO                & Defender agents compete against a changing team of heuristic attackers to protect network nodes.                                \\
    Ride-Sharing            & TO                & Agents cooperatively and competitively deliver passengers to maximize individual fares while minimizing passenger waiting time. \\
    \bottomrule
  \end{tabular}
  \label{tab:tracks_2026}
\end{table*}

\subsection{Differences from the 2025 Competition}

The 2026 competition differed from the inaugural competition in two primary ways.

\textbf{Frame openness in wildfire.} The 2026 competition introduced frame openness in a bonus wildfire track. In this track, agents may have changing equipment states, including suppressant capacities and firefighting ranges. These equipment changes affect how long agents can remain effective in the environment and which fires they can reach, requiring policies to reason not only about agent and task availability but also about changes in agent capabilities.

\textbf{Task-centered metrics.} The 2026 evaluation placed greater emphasis on interpretable task-level outcomes. In addition to domain rewards, the competition introduced or highlighted metrics such as total task completions, mean task-completion time, and mean value of completed tasks. For ride-sharing, these correspond naturally to the number of completed passenger trips, passenger waiting time, passenger ride time, and fare or reward-related values.

\subsection{Participation and Submission Summary}

Eight teams registered for the 2026 competition. Of the four available tracks, only the ride-sharing track received a final submission. The submitted team, DLC, is therefore recognized as the winner of the 2026 MOASEI ride-sharing track. Table~\ref{tab:submission_summary} summarizes the submission.

\begin{table}[htbp]
  \centering
  \caption{Submitted MOASEI 2026 competition approach.}
  \begin{tabular}{lll}
    \toprule
    \textbf{Team} & \textbf{Approach}         & \textbf{Domain} \\
    \midrule
    DLC           & Planning with replanning  & Ride-sharing    \\
    \bottomrule
  \end{tabular}
  \label{tab:submission_summary}
\end{table}

The DLC policy uses a planning approach that solves an optimal routing problem across agents. It replans when new passengers enter the environment, enabling the policy to adapt to task openness in the ride-sharing domain.

\section{Evaluation Methodology and Results Analysis}
\label{sec:eval_methods_and_results}

Because the only final submission was for ride-sharing, the 2026 results analysis focuses on that track. For readers interested in prior results for the wildfire fighting and cybersecurity tracks, please consult our technical report for the 2025 competition~\cite{patino2025_inaugural}.  

The ride-sharing environment evaluates agents that accept, pick up, and drop off passengers that appear over time. We report three primary metrics from the evaluation logs: completed passengers, average passenger wait time, and average passenger ride time. Completed passengers measures task throughput, while wait and ride times characterize the quality and efficiency of service.

The submitted DLC policy was evaluated on six ride-sharing configurations (\texttt{rideshare\_0} through \texttt{rideshare\_5}) across 213 matched evaluation episodes. These six configurations (three provided to contestents for training, three held back for testing evaluation) vary in the starting positions of ride-sharing driver agents, the total numbers of passengers, and the schedules of passenger arrival in the system.  Together, these configurations provide a range of scenarios and task sequences within which agents must make challenging decisions.  

For comparison and richer evaulation of the DLC submission, we also report baseline policies including: no-operation, random, first-in-first-out (FIFO) variants, and greedy variants. These baselines are not competition submissions, but they provide useful context for interpreting the submitted policy.

\subsection{Overall Ride-shareing Results}

Table~\ref{tab:rideshare_overall} shows the aggregate performance of DLC and the baseline policies.  Relative to the baseline policies, DLC completed substantially more passengers than FIFO, random, and no-operation policies while maintaining competitive passenger wait time, and successfully completed nearly as many rides as the strongest greedy baseline. For the no-operation policy, wait and ride times are reported as zero because no passengers were completed in the evaluation logs.

\begin{table*}[htbp]
  \centering
  \caption{Ride-sharing results across all configurations. Values are means $\pm$ sample standard deviations over 213 episodes. Baseline policies are included for context and were not submitted competition teams.}
  \begin{tabular}{lccc}
    \toprule
    \textbf{Method} & \textbf{Completed Passengers} & \textbf{Avg. Wait Time} & \textbf{Avg. Ride Time} \\
    \midrule
    dlc & $13.62 \pm 0.88$ & $24.13 \pm 4.51$ & $34.87 \pm 4.04$ \\
    fifo\_Tfocus & $10.80 \pm 0.89$ & $39.72 \pm 3.64$ & $14.21 \pm 1.68$ \\
    fifo\_Tglobal & $10.07 \pm 1.14$ & $36.75 \pm 4.77$ & $17.91 \pm 4.99$ \\
    greedy\_Tfocus & $14.54 \pm 1.30$ & $30.04 \pm 3.31$ & $12.07 \pm 0.99$ \\
    greedy\_Tglobal & $15.38 \pm 1.60$ & $25.48 \pm 3.64$ & $12.14 \pm 2.07$ \\
    noop & $0.00 \pm 0.00$ & $0.00 \pm 0.00$ & $0.00 \pm 0.00$ \\
    random & $7.40 \pm 2.30$ & $31.28 \pm 9.24$ & $48.55 \pm 10.38$ \\
    \bottomrule
  \end{tabular}
  \label{tab:rideshare_overall}
\end{table*}

\begin{figure}[htbp]
  \centering
  \includegraphics[width=\columnwidth]{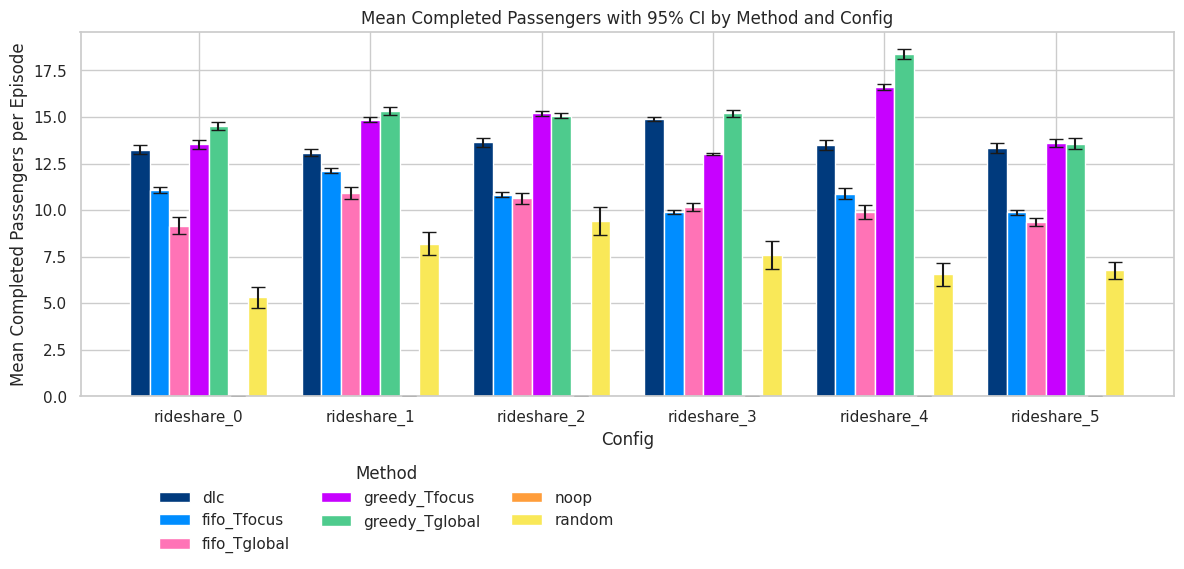}
  \Description{Bar chart comparing the mean number of completed passengers across the DLC policy and ride-sharing baselines.}
  \caption{Mean completed passengers for ride-sharing policies.}
  \label{fig:rideshare_completed_passengers_mean}
\end{figure}

\begin{figure}[htbp]
  \centering
  \includegraphics[width=\columnwidth]{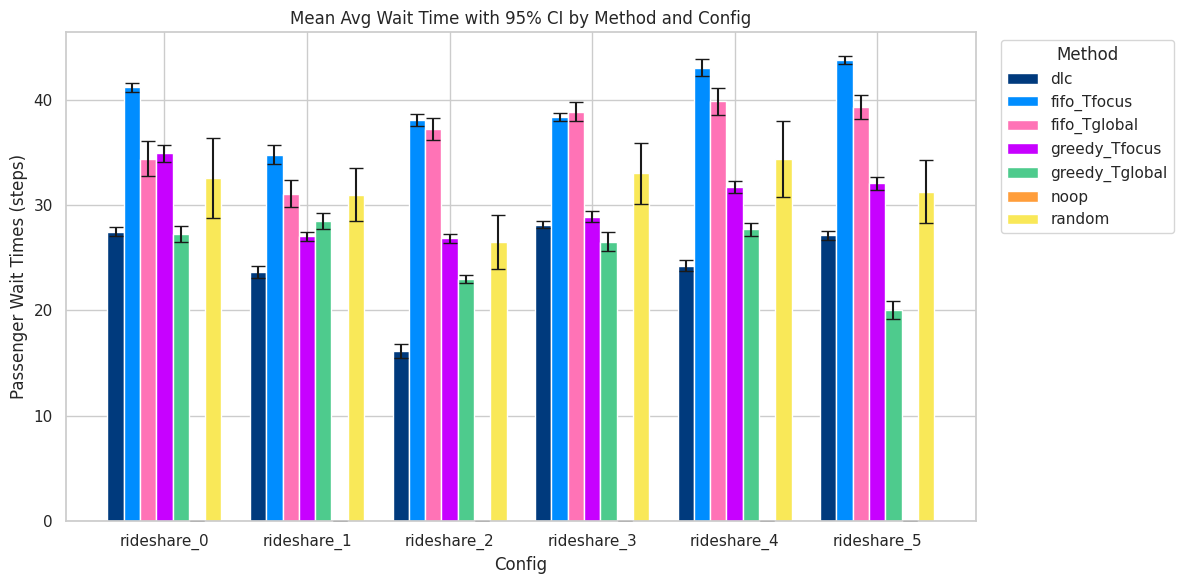}
  \Description{Bar chart comparing mean passenger wait time across the DLC policy and ride-sharing baselines.}
  \caption{Mean passenger wait time for ride-sharing policies.}
  \label{fig:rideshare_wait_time_mean}
\end{figure}

\begin{figure}[htbp]
  \centering
  \includegraphics[width=\columnwidth]{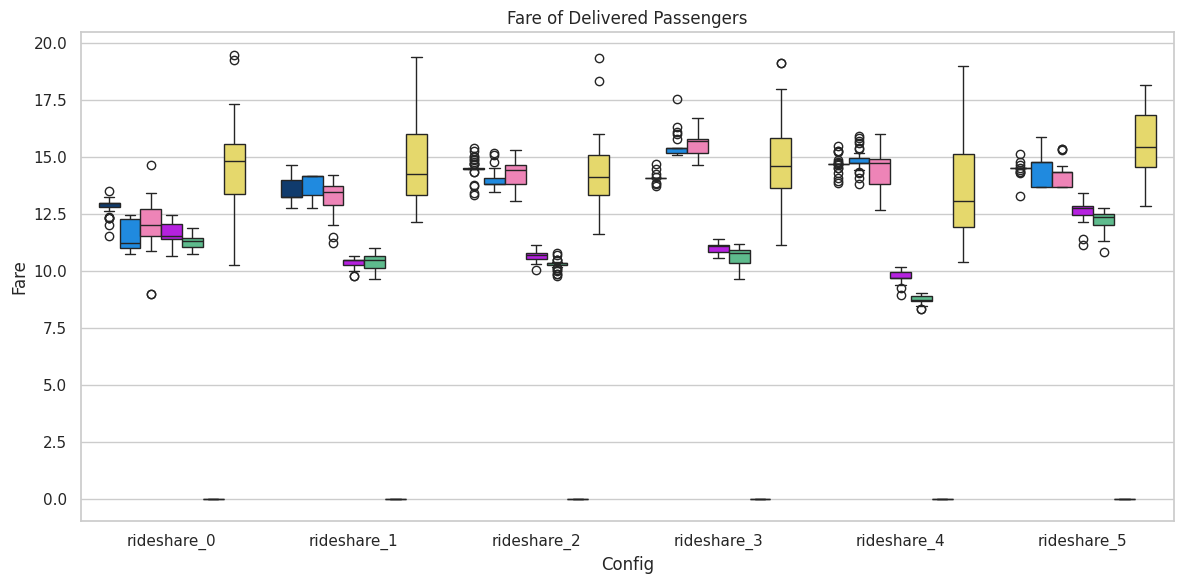}
  \Description{Bar chart comparing mean fare value for completed tasks across ride-sharing policies.}
  \caption{Mean fare value for completed ride-sharing tasks.}
  \label{fig:rideshare_fare_value_mean}
\end{figure}

\begin{figure}[htbp]
  \centering
  \includegraphics[width=\columnwidth]{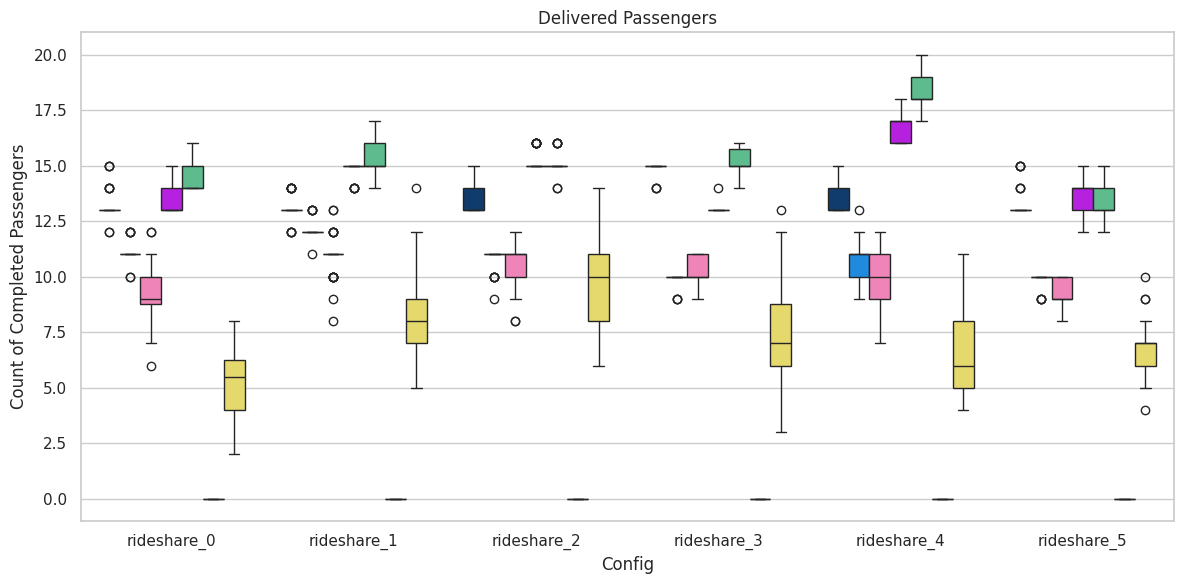}
  \Description{Configuration-level plot of completed passengers for the DLC policy and ride-sharing baselines.}
  \caption{Completed passengers by ride-sharing configuration.}
  \label{fig:rideshare_completed_passengers_config}
\end{figure}

\begin{figure}[htbp]
  \centering
  \includegraphics[width=\columnwidth]{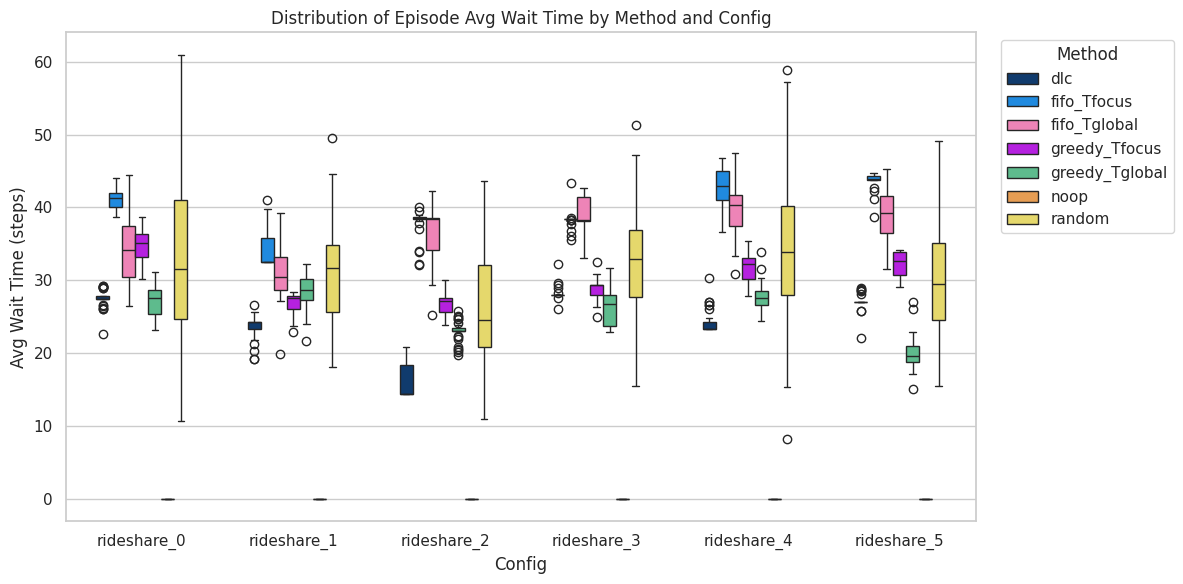}
  \Description{Configuration-level plot of passenger wait time for the DLC policy and ride-sharing baselines.}
  \caption{Passenger wait time by ride-sharing configuration.}
  \label{fig:rideshare_wait_time_config}
\end{figure}

\begin{figure}[htbp]
  \centering
  \includegraphics[width=\columnwidth]{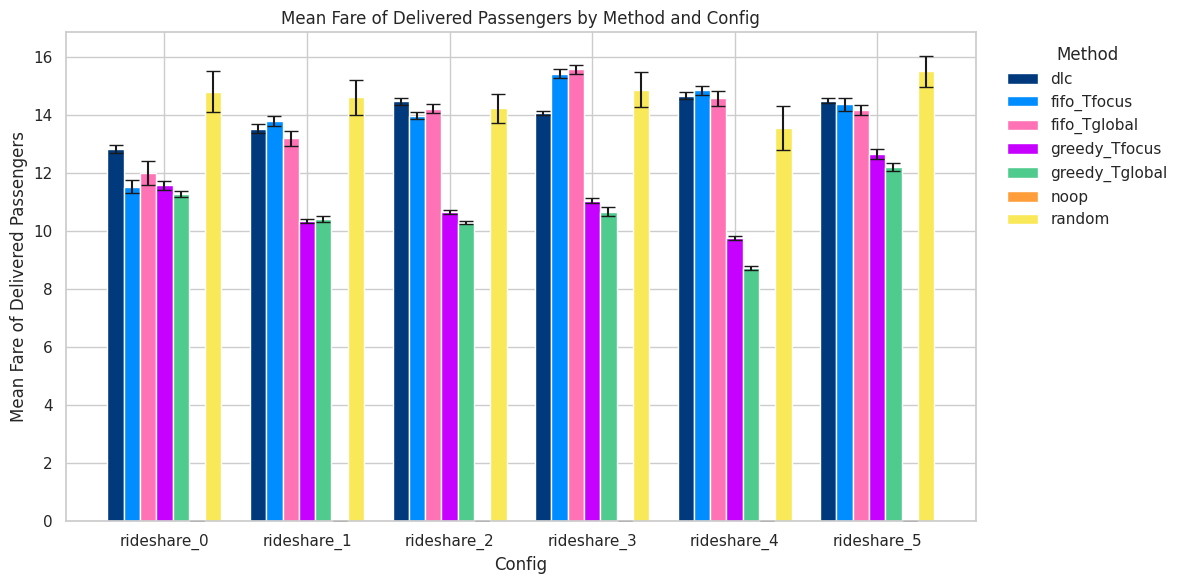}
  \Description{Configuration-level plot of fare value for the DLC policy and ride-sharing baselines.}
  \caption{Fare value by ride-sharing configuration.}
  \label{fig:rideshare_fare_value_config}
\end{figure}

\begin{figure}[htbp]
  \centering
  \includegraphics[width=\columnwidth]{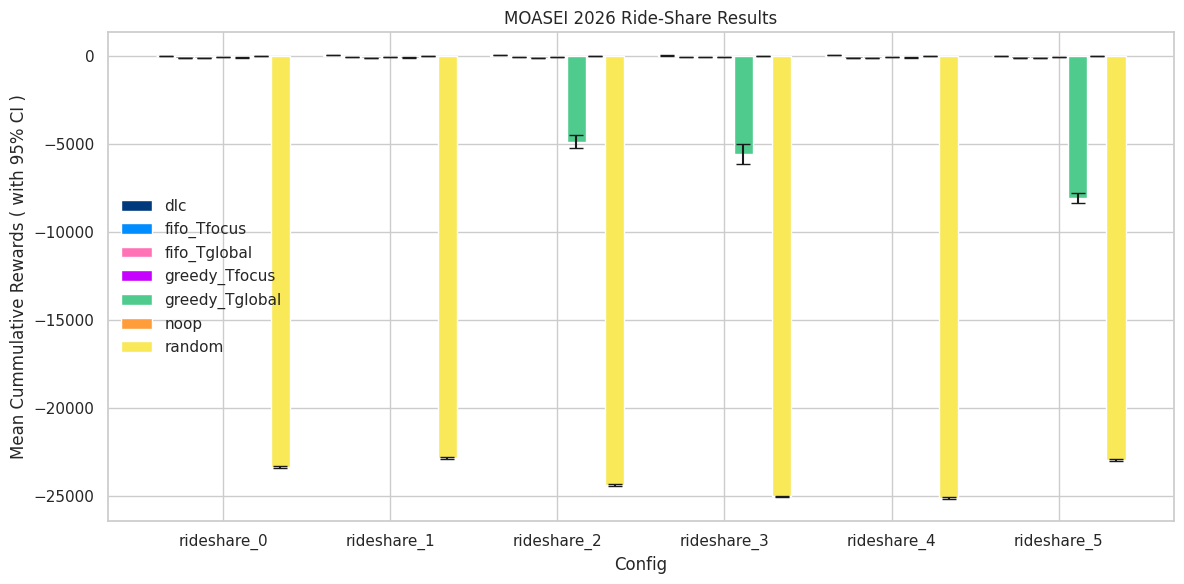}
  \Description{Plot comparing mean cumulative reward across ride-sharing policies.}
  \caption{Mean cumulative reward for ride-sharing policies.}
  \label{fig:rideshare_rewards_cumulative_mean}
\end{figure}

\begin{figure}[htbp]
  \centering
  \includegraphics[width=\columnwidth]{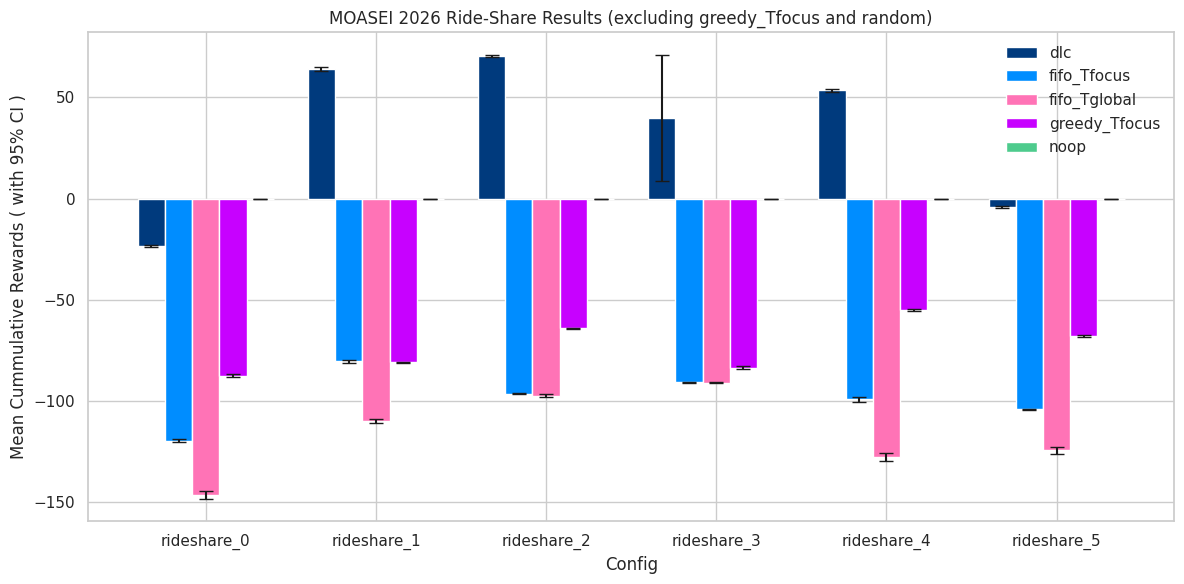}
  \Description{Plot comparing mean cumulative reward after removing selected high-variance or dominant baselines to make the remaining policies easier to inspect.}
  \caption{Mean cumulative reward with selected baselines removed for readability.}
  \label{fig:rideshare_rewards_cumulative_zoomed}
\end{figure}

Figure~\ref{fig:rideshare_completed_passengers_mean} visualizes the mean number of completed passengers across methods. Figure~\ref{fig:rideshare_wait_time_mean} shows corresponding mean passenger wait times. Figure~\ref{fig:rideshare_fare_value_mean} summarizes mean fare value, which provides the ride-sharing specific view of the value of completed tasks. We also include configuration-level and cumulative-reward plots in Figures~\ref{fig:rideshare_completed_passengers_config}-~\ref{fig:rideshare_rewards_cumulative_zoomed} to provide additional  information about the task-level behavior of agents.

\subsection{DLC Performance by Configuration}

Table~\ref{tab:dlc_by_config} reports DLC performance for each ride-sharing evaluation configuration. DLC completed between 13.08 and 14.91 passengers on average across the six configurations. Its lowest wait time occurred in \texttt{rideshare\_2}, while its highest average throughput occurred in \texttt{rideshare\_3}. Ride time varied more than throughput, indicating that the policy's routing decisions were stable in the number of trips completed but sensitive to configuration-level spatial and task-arrival differences.

\begin{table}[htbp]
  \centering
  \caption{DLC ride-sharing performance by configuration. Values are means $\pm$ sample standard deviations.}
  \begin{tabular}{lccc}
    \toprule
    \textbf{Config} & \textbf{Completed} & \textbf{Wait}    & \textbf{Ride}    \\
    \midrule
    rideshare\_0    & $13.25 \pm 0.72$   & $27.48 \pm 1.27$ & $33.32 \pm 2.44$ \\
    rideshare\_1    & $13.08 \pm 0.55$   & $23.65 \pm 1.60$ & $32.94 \pm 2.44$ \\
    rideshare\_2    & $13.63 \pm 0.80$   & $16.12 \pm 2.11$ & $38.68 \pm 2.02$ \\
    rideshare\_3    & $14.91 \pm 0.29$   & $28.12 \pm 0.90$ & $30.48 \pm 2.95$ \\
    rideshare\_4    & $13.50 \pm 0.74$   & $24.25 \pm 1.54$ & $33.64 \pm 1.75$ \\
    rideshare\_5    & $13.33 \pm 0.69$   & $27.13 \pm 1.16$ & $39.65 \pm 2.65$ \\
    \bottomrule
  \end{tabular}
  \label{tab:dlc_by_config}
\end{table}

\FloatBarrier

\section{Findings from the MOASEI Competition}

The 2026 competition produced two main findings. First, the ride-sharing submission demonstrated that planning-based replanning is a viable approach for task-open environments. DLC adapted to newly arriving passengers and achieved stable passenger-completion counts across all six evaluation configurations. Its average wait time was substantially lower than FIFO baselines and close to the stronger greedy baseline, suggesting that replanning helped the policy respond to passenger arrivals in a timely manner.

Second, the baseline comparison highlights an important objective tradeoff. DLC achieved strong throughput and wait-time performance, but its ride times were longer than the greedy and FIFO baselines. This suggests that the policy may prioritize accepting and completing more passenger tasks even when the resulting routes are longer. Future versions of the ride-sharing evaluation should make the relative importance of completions, waiting time, ride time, fare value, and cumulative reward explicit in the leaderboard objective.

The limited number of final submissions also shaped the interpretation of the results. Unlike the 2025 competition, the 2026 event does not support broad conclusions about the relative strengths of different learning or planning paradigms. The main empirical contribution of this year's competition is therefore a focused case study of one planning-based ride-sharing policy and a record of the updated benchmark design.

\section{Future Work}

Future MOASEI competitions should focus on improving participation while preserving the richer evaluation structure introduced in 2026. The addition of frame openness in the bonus wildfire track provides a useful direction for future benchmark design, but the track will require clearer examples, starter policies, and documentation to encourage submissions. Similarly, the ride-sharing track would benefit from a published leaderboard objective that balances passenger completions, waiting time, ride time, and fare value.

We also plan to continue developing baseline policies and public evaluation artifacts. Strong baselines are useful for interpretation, but they should be clearly separated from submitted competition entries. Publishing baseline implementations, evaluation seeds, and result summaries will make future comparisons more transparent and reproducible.

\section{Conclusion}
\balance

The second MOASEI Competition continued the initiative's goal of evaluating multi-agent systems under open-world conditions, while introducing frame openness and more task-centered metrics. Although only one of four tracks received a final submission, the ride-sharing results provide a useful benchmark for planning under task openness. Notably, these results fill a gap from the first MOASEI competition~\cite{patino2025_inaugural}, which only received submissions for the wildfire fighting and cybersecurity tracks.  Thus, the two competitions together provide a more comprehensive view of multiagent decision making in open agent systems.

DLC is recognized as the 2026 winner of the ride-sharing track. Its planning and replanning approach achieved stable passenger completions and competitive wait times, while also revealing tradeoffs in ride duration. These results motivate clearer objective design, stronger onboarding materials, and continued development of openness-focused evaluation tracks for future MOASEI competitions.

\section{Acknowledgements} 
This research was supported, in part, by a collaborative NSF Grant \#IIS-2312657 (to Doshi), \#IIS-2312658 (to Soh), and \#IIS-2312659 (to Eck). Some of the computing occurred on the Holland Computing Center of the University of Nebraska, which receives support from the university's Office of Research and Economic Development and the Nebraska Research Initiative. 

\bibliographystyle{ACM-Reference-Format} 
\bibliography{references}

\end{document}